\documentclass{elsarticle}
\usepackage{amsmath,amssymb,epsfig}
\usepackage[backref,citecolor=blue]{hyperref}
\def\epspdffile#1{\leavevmode\ifpdf\epsffile{#1.pdf}\else\epsffile{#1.eps}\fi}
\def\Dslash{\mathchoice
    {D\hskip-0.62em\raise0.2ex\hbox{$\displaystyle/$}\hskip0.2em}%
    {D\hskip-0.62em\raise0.2ex\hbox{$\textstyle/$}\hskip0.2em}%
    {D\hskip-0.5em\raise0.15ex\hbox{$\scriptstyle/$}\hskip0.2em}%
    {D\hskip-0.5em\raise0.15ex\hbox{$\scriptscriptstyle/$}\hskip0.2em}}
\def\dslash{\mathchoice
    {\partial\hskip-0.5em\raise0.2ex\hbox{$\displaystyle/$}\hskip0.2em}%
    {\partial\hskip-0.5em\raise0.2ex\hbox{$\textstyle/$}\hskip0.2em}%
    {\partial\hskip-0.4em\raise0.15ex\hbox{$\scriptstyle/$}\hskip0.2em}%
    {\partial\hskip-0.4em\raise0.15ex\hbox{$\scriptscriptstyle/$}\hskip0.2em}}
\def\Aslash{\mathchoice
    {A\hskip-0.5em\raise0.2ex\hbox{$\displaystyle/$}\hskip0.2em}%
    {A\hskip-0.5em\raise0.2ex\hbox{$\textstyle/$}\hskip0.2em}%
    {A\hskip-0.4em\raise0.15ex\hbox{$\scriptstyle/$}\hskip0.2em}%
    {A\hskip-0.4em\raise0.15ex\hbox{$\scriptscriptstyle/$}\hskip0.2em}}
\def\gwop{\tilde{\Dslash}}          
\def\nop#1{\Dslash_{#1}}            
\def\nopex#1{\tilde{\Dslash}_{#1}}  
\def\lowbound{\xi}                  
\def\wop{\Dslash_W}                 
\def\zop{\Dslash_Z}		    
\def\hop{\Dslash_{\mbox{\tiny KLN}}} 
\def\zdeg{N}                         
\def\sgn{\mathop{\rm sgn}\nolimits}  
\def\appsgn{\varepsilon}             
\def\zsgn{\appsgn_{\mbox{\tiny Z}}}  
\def\hsgn{\appsgn_{\mbox{\tiny KLN}}} 
\def\defect{\Delta}                   
\def\Re{\mathop{\rm Re}}              
\def\tr{\mathop{\rm tr}}              
\def\sn{\mathop{\rm sn}}              
\def\K{\mathop{\rm K}}                
\def\sameas{\quad\Leftrightarrow\quad} 
\def\defn{\equiv}                      
\def\mres{m_{\mbox{\tiny res}}}        
\def\dt{\delta\tau}                    
\newif\ifdraft \draftfalse           
\def\note[#1]#2{\message{(#1)}\ifdraft{\noindent\bf[#2]\/}\fi}
\ifdraft
\fi
\def\rational#1#2{{\mathchoice{\textstyle{\frac{#1}{#2}}}%
  {\scriptstyle{\frac{#1}{#2}}}{\scriptscriptstyle{\frac{#1}{#2}}}{#1/#2}}}
\def\half{\rational12}		    
\begin{document}
\begin{frontmatter}
  \title{A lattice Dirac operator\\ for QCD with light dynamical quarks}
  \author[a,c]{Nigel Cundy}
  \author[b]{A D Kennedy}
  \author[a]{Andreas Sch\"afer}
  \address[a]{Institut f\"ur Theoretische Physik, Universit\"at Regensburg,
    D-93040 Regensburg, Germany}
  \address[b]{SUPA, School of Physics \& Astronomy,\\ University of Edinburgh,
    Edinburgh EH9~3JZ, Scotland}
  \address[c]{Lattice Gauge Theory Research Center, FPRD, and CTP,\\ Department of Physics \&
    Astronomy, Seoul National University, Seoul, 151-747,\\ South Korea}
  \begin{abstract}
    In QCD chiral symmetry is explicitly broken by quark masses, the effect of
    which can be described reliably by chiral perturbation theory. Effects of
    explicit chiral symmetry breaking by the lattice regularisation of the
    Dirac operator, typically parametrised by the residual mass, should be
    negligible for almost all observables if the residual mass of the Dirac
    operator is much smaller than the quark mass.  However, maintaining a
    small residual mass becomes increasingly expensive as the quark mass
    decreases towards the physical value and the continuum limit is
    approached.  We investigate the feasibility of using a new approximately
    chiral Dirac operator with a small residual mass as an alternative to
    overlap and domain wall fermions for lattice simulations.  Our Dirac
    operator is constructed from a Zolotarev rational approximation for the
    matrix sign function that is optimal for bulk modes of the Hermitian
    kernel Dirac operator but not for the low-lying parts of its spectrum.  We
    test our operator on various \(32^3\times64\) lattices, comparing the
    residual mass and the performance of the Hybrid Monte Carlo algorithm at a
    similar lattice spacing and pion mass with a hyperbolic tangent operator
    as used by domain wall fermions.  We find that our approximations have a
    significantly smaller residual mass than domain wall fermions at a similar
    computational cost, and still admit topological charge change.
  \end{abstract}
\end{frontmatter}

\ifdraft
  \begin{center} \fbox{\ifpdf PDF\fi La\TeX\ version of \today.} \end{center}
\fi

\section{Introduction}
Lattice QCD is a phenomenologically successful regularisation of QCD which can
accurately predict experimental observables. It places space-time on a
discrete lattice, and the continuum theory is recovered as three limits are
taken, the volume to infinity, the lattice spacing to zero, and the quark
masses to their physical values. The extrapolations in the lattice spacing and
quark mass are made more difficult, with increased errors, by explicit
breaking of chiral symmetry within the lattice discretization of the Dirac
operator. The Nielson-Ninomiya theorem~\cite{Nishy-Ninny}, which states that
it is impossible to represent an odd number of quark flavours on the lattice
without breaking chiral symmetry is avoided for the cheapest commonly used
families of lattice actions either by simulating additional degenerate
fermions (staggered fermions) or by explicitly breaking chiral symmetry
(Wilson fermions).

There is, however, an alternative, to use a lattice chiral
symmetry~\cite{Luscher:1998pqa}, defined by a Ginsparg-Wilson
relation~\cite{Ginsparg:1982bj}, which has a smooth limit to the continuum
chiral symmetry. The only known and practical lattice Dirac operators which
satisfy lattice chiral symmetry are discontinuous, being built from the matrix
sign function. The simplest of these is the overlap operator, first proposed
by Neuberger and Narayanan~\cite{Narayanan:1993sk,Neuberger:1998my}. However,
the computational difficulties involved in simulating overlap fermions mean
that their use in lattice simulations remains challenging. A previous solution
with a a good approximate chiral symmetry, the domain wall fermion, which
placed left handed and right handed fermions on the four dimensional surfaces
at the boundaries of a five dimensional lattice~\cite{Kaplan:1992bt}. This
formalism reduces to a form of the overlap operator for infinite fifth
dimension, giving exactly chiral fermions. It is also equivalent to the four
dimensional Kenney-Laub-Neuberger \cite{kenney:1994,Neuberger:1998my} Dirac
operator.

There are, however, many other ways in which an approximate chiral symmetry
can be maintained on the lattice. There are no \textit{a priori} reasons to
suspect that the historical domain wall fermion should be the best (the `best'
implementation can be defined as the one that requires the least computational
effort for a particular residual mass, the standard measurement of explicit
chiral symmetry breaking on the lattice). In this work, we propose and
investigate a four dimensional approximation to the overlap operator that
gives a considerably better residual mass than domain wall fermions while
avoiding the complexities of exact overlap fermions.

\subsection{On-Shell Chiral Symmetry}

For a lattice Dirac operator \(\gwop\), an on-shell chiral transformation can
be written as~\cite{Luscher:1998pqa} \[\psi\mapsto
e^{i\alpha\gamma_5(1-a\gwop)}\psi
\qquad\mbox{and}\qquad\bar\psi\mapsto\bar\psi e^{i\alpha(1-a\gwop)\gamma_5};\]
the differential condition that the Dirac operator itself is invariant under
this transformation is a Ginsparg--Wilson relation~\cite{Ginsparg:1982bj}
\[\{\gamma_5,\gwop\} = 2a\gwop\gamma_5\gwop,\] which may be written in
the equivalent forms \[\Re\gwop = a\gwop^\dagger\gwop \sameas \Re[(1-
a\gwop)^\dagger a\gwop]=0 \sameas \Re\gwop^{-1} = a.\] We may define the
quantity \(\hat\gamma_5\) by \(a\gwop = \half(1+\gamma_5 \hat \gamma_5)\), and
if we wish \(\gwop\) to satisfy \(\gwop^\dagger = \gamma_5\gwop \gamma_5\)
then \(\hat\gamma_5 =\hat\gamma_5^\dagger\).  This Ginsparg--Wilson relation
implies that \(\hat\gamma_5^2=1\), so \(\hat \gamma_5\) is both hermitian and
unitary, so that its spectrum can contain only the eigenvalues \(\pm1\) and
hence \(\hat\gamma_5 = \sgn(\hat\gamma_5)\).

We must also require that \(\gwop\) is a Dirac operator in the na{\"\i}ve
continuum limit, \(\gwop = Z_{GW}(\dslash+\Aslash) + O(a^2)\), where there are
no corrections of \(O(a)\) because the only available such operator is
\([\dslash+\Aslash,\dslash+\Aslash] = \sigma\cdot F \defn \sigma_{\alpha\beta}
F^{\alpha\beta}\) which is not chirally invariant.  If \(\wop\) is any
\(\gamma_5\)-hermitian lattice Dirac operator satisfying \(\wop =
Z_W(\dslash+\Aslash) + O(a)\) then we may choose \(2aM\gwop = \wop + O(a)\)
where \(2aM = Z_W/Z_{GW}\) is a suitable finite wave-function renormalization,
and we obtain the (massless) Neuberger
operator~\cite{Neuberger:1998fp,Neuberger:1998wv}
\begin{equation}
a\nopex0 \defn \half
(1+\gamma_5\sgn H)\label{eq:masslessNeuberger}
\end{equation}
 where \(H\defn\gamma_5\wop-M\).  In QCD a quark has a
(small) mass \(\mu\), and this can be incorporated by shifting the spectrum to
lie in the interval \([a\mu,1]\) by defining the massive Neuberger operator to
be
\begin{equation}
  a\gwop\mu \defn \half[(1+a\mu) + (1-a\mu)\gamma_5 \sgn(H)]. 
  \label{eq:Neuberger}
\end{equation}

Within this framework there are numerous choices that have to be
made~\cite{Kennedy:2006ax}: whether to use four or five dimensional
pseudofermions; the choice of the kernel operator \(H\) (including the mass
\(M\)); the choice of rational approximation for the matrix sign function;
which form of 5D matrix with the desired Schur complement (for example,
Euclidean-Caley or continued fraction); and for four dimensional
pseudofermions whether to use a nested 4D or 5D inverter. All of these choices
are independent and physically equivalent. We use the name ``domain wall
operator'' as any approach which uses five dimensional pseudofermions, and
``overlap operator'' as any approach in four dimensions with an exact matrix
sign function, limited only by the floating point precision of the
computer. An ``approximate overlap operator'' is a four dimensional approach
which has a small explicit breaking of chiral symmetry from the use of an
approximate matrix sign function. In this article, we are only studying the
effect of changing the rational approximation used for the matrix sign
function. Our results should be independent of the choice of kernel and
whether four dimensional or five dimensional pseudofermions are used.
\subsection{Ginsparg--Wilson Defect and Residual Mass}
\label{sec:residual-mass}

Suppose we have some approximation \(\appsgn(H)\approx\sgn(H)\) to the matrix
sign function, so that equation (\ref{eq:Neuberger}) is replaced by
\begin{equation}
D_{\mu} = \frac{1}{2}[(1+a\mu) + (1-a\mu)\appsgn(H)],\label{eq:Neuberger2}
\end{equation}
then we may define the defect \(\defect\) as the amount by which the
corresponding approximate Neuberger operator, \(\nop{} \), fails to satisfy
the Ginsparg--Wilson relation for \(\mu=0\), \[\defect \defn \half\{\gamma_5,
\nop0\}-a\nop0\gamma_5\nop0 = \gamma_5\Re[(1-a\nop0)^\dagger \nop0],\] which
gives
\begin{equation}
  4\gamma_5 a\defect = 1-\appsgn(H)^2. \label{eq:defect}
\end{equation}
From equation~(\ref{eq:Neuberger2}) we have \(\nop\mu = \nop0 + \mu(1-
a\nop0)\), hence
\begin{eqnarray*}
  \Re[(1-a\nop0)^\dagger \nop\mu]
    &=& \Re[(1-a\nop0)^\dagger \nop0] +
      \mu(1-a\nop0)^\dagger(1-a\nop0) \\
    &=& \gamma_5 \defect + \mu(1-a\nop0)^\dagger(1-a\nop0).
\end{eqnarray*}
Multiplying this by \({\nop\mu^\dagger}^{-1}\) on the left and
\(\nop\mu^{-1}\) on the right we obtain
\begin{equation}
  \half(S_\mu+S_\mu^\dagger) 
    = {\nop\mu^\dagger}^{-1} \gamma_5\defect\nop\mu^{-1} 
      + \mu S_\mu^\dagger S_\mu;  
  \label{eq:res-mu}
\end{equation}
where \(S_\mu\defn(1-a\nop0)\nop\mu^{-1}\) (which satisfies \(S_\mu^{-1} =
S_0^{-1} +\mu\)).  The first term on the right side of
equation~(\ref{eq:res-mu}) is a measure of the chiral symmetry breaking due to
the approximation to the sign function whereas the second is that due to the
explicit quark mass~\(a\mu\).  We wish to introduce some norm \(\|\defect\|\)
on the defect to quantify the magnitude of the errors due to our approximation
to the sign function.  A useful estimate for this norm is the residual mass
\[\mres'\defn\frac{\tr({\nop\mu^\dagger}^{-1}\gamma_5\defect\nop\mu^{-1})}
{\tr(S_\mu^\dagger S_\mu)},\] whence \(\tr(S_\mu+S_\mu^\dagger)/2\tr(S_\mu
S_\mu^\dagger) = \mres + \mu\).

There is no a priori reason when comparing the extent of chiral symmetry
breaking for different Dirac operators why we should project the defect onto a
scalar using this trace, indeed in our numerical studies we have used the
computationally cheaper approach of projecting onto momentum zero states,
giving \[\mres = \frac{\sum_{x,x'} \left[{\nop\mu^\dagger}^{-1}
\gamma_5\defect \nop\mu^{-1} \right]_{x,x'}}{\sum_{x,x'} \left[S_\mu^\dagger
S_\mu \right]_{x,x'}}.\] We do not expect that the quantity \(\mres\) will be
less suitable than ~\(\mres'\) to compare the effects of chiral symmetry
breaking between different Dirac operators.

Since our approximation for the Neuberger operator does not exactly satisfy
chiral symmetry it will have \(O(a)\) corrections near the continuum limit,
and as these must either come from the explicit mass term or the defect we see
that
\begin{equation}
  \nop\mu = Z_{GW}(\dslash+\Aslash)
    + \mu + \mres+\mbox{constant}\times a\sigma\cdot F + O(a^2).
  \label{eq:continuum-limit}
\end{equation}

\subsection{Effects of Residual Mass}

The spectrum of the exact unitary matrix \(\gamma_5\hat\gamma_5 = \gamma_5
\sgn(H)\) lies on the unit circle in the complex plane \(\|\gamma_5
\hat\gamma_5\| = 1\).  The approximate matrix sign function has spectral norm
\(\|\appsgn(H)\|\defn\sup_{|\psi|=1} \psi^\dagger \appsgn(H)\psi\), so the
spectrum of \(\|\gamma_5\appsgn(H)\| \leq \|\gamma_5\|\,\|\appsgn(H)\| =
\|\appsgn(H)\|\) lies within a disc of radius \(\|\appsgn(H)\|\).  For an
operator with good chiral symmetry most eigenvalues will be close to the unit
circle; however small eigenvalues of \(H\), where the approximation is less
good, may exhibit large discrepancies from the spectrum of the exact overlap
operator.  For example, if \(H\) has an exactly zero eigenvalue then for all
symmetric approximations to the matrix sign function the corresponding
eigenvalue of \(\nop\mu\) will lie in the centre of the circle, i.e. at
\(\half(1+a\mu)\).

If the residual mass is sufficiently smaller than the target physical quark
mass, \(\mres\ll\mu\) then as long as the lattice spacing is small enough that
the \(O(a^2)\) effects in equation~(\ref{eq:continuum-limit}) are negligible
all physical effects of \(\mres\) can be removed by adjusting \(\mu\mapsto
\mu-\mres\) so that the quark mass stays fixed.  However, if the residual mass
is larger than the physical quark mass then this is not possible in general.
While one could insert \(\mu<0\) into the lattice Dirac operator this would be
likely to introduce zero or negative eigenvalues, leading to the inexact
overlap Dirac operator becoming singular on some (now exceptional)
configurations.

It is therefore desirable to have a small residual mass.  The difficulty with
this is that equation~(\ref{eq:defect}) tells us that \(\|\defect\| =
\|1-\appsgn(H)\|/4a\), so we need to reduce the error \(\|1-\appsgn(H)\|
\propto a\) in our approximation to keep the physical \(\mres\) fixed as the
continuum limit is approached.  This can be done by increasing the order of
the rational approximation which may, depending on the approximation used,
significantly increase the cost of the simulation.

At larger lattice spacings it is easier to maintain a small \(\mres\), but
there will be larger \(O(a^2)\) and higher lattice artefacts in
equation~(\ref{eq:continuum-limit}); since these come from the chiral symmetry
breaking in \(\defect\) and not just from the quark mass \(\mu\) they are less
easy to model using, for example, chiral perturbation theory.

\subsection{Choice of Rational Approximation}

We have shown that we need a lattice Dirac operator with a good approximation
to (on-shell) chiral symmetry, but we also wish to avoid the cost of
maintaining exact chiral symmetry to machine precision.  Our goal is thus to
find a family of approximations to the matrix sign function that provides a
good balance between residual mass and the cost of Hybrid Monte Carlo (HMC)
computations.  We shall demonstrate that domain wall fermions, where the
accuracy of whose hyperbolic tangent approximation falls too slowly with
increasing rational degree are far from optimal, while overlap fermions, which
are far more chiral than necessary for massive quarks at a cost in time and
complexity of the HMC algorithm, are more expensive than is required to obtain
a good enough chiral symmetry.

In this work, we introduce and test the Zolotarev lattice Dirac operator,
which uses the optimal rational approximation to the matrix sign function, but
not over the entire spectrum of the kernel operator \(\wop\).  This is
guaranteed to provide the best approximation to the matrix sign function for a
given order of rational approximation within a certain tunable eigenvalue
range, and therefore might be expected to give the smallest residual mass for
a given amount of computational effort.  However, there are a number of
questions which need to be addressed in this comparison.
\begin{enumerate}
\item While the Zolotarev approximation is guaranteed to give the best
  \(L_\infty\) approximation over part of the spectrum, some eigenvalues lie
  outside this range. What contribution do they make to the residual mass?
\item Does the reduced residual mass come at the cost of less stable molecular
  dynamics? If a small time step were required to resolve larger fluctuations
  in the fermionic force this could compensate for any gain in the residual
  mass.
\item Is the Zolotarev Dirac operator ergodic? In particular, are all
  topological sectors sampled, and is the autocorrelation between topological
  sectors and different portions of the same topological sectors short enough?
\item Is the Zolotarev Dirac operator local?  As it does not approximate the
  sign function well for small eigenvalues of the kernel operator it is not
  obvious that we can rely on previous proofs and numerical results, although
  it seems unlikely that it could be less local than the hyperbolic tangent
  Dirac operator used in domain wall computations.
\end{enumerate}

In \S\ref{sec:zolotarev} we describe our Zolotarev approximation and compare
it to the Kenney--Laub--Neuberger \cite{kenney:1994,Neuberger:1998my}
hyperbolic tangent approximation, which is a four dimensional representation
of the five dimensional domain wall Dirac operator; in
\S\ref{sec:numerical-implementation} we describe how we tested the various
methods; in \S\ref{sec:results} we present the numerical results from these
tests; and in \S\ref{sec:conclusions} we present our conclusions.

\section{The Zolotarev Dirac operator} \label{sec:zolotarev}

\subsection{The Zolotarev approximation}

The Zolotarev Dirac operator depends on the degree \(\zdeg\) of the rational
approximation and a parameter \(\lowbound\) where \(\lowbound\|H\|\leq
|\lambda|\leq \|H\|\) is the interval of the kernel operator's spectrum on
which the approximation to the matrix sign function is optimal.  Here
\(\displaystyle \|H\|=\sup_{\|u\|_2= 1}(u,Hu)\) is the spectral norm of the
hermitian kernel operator \(H\), i.e., its largest eigenvalue.  The Zolotarev
Dirac operator is \[a\zop = \half\left[(1+a\mu) + (1-a\mu)\gamma_5\,
\zsgn\!\left(\frac H{\|H\|} \right)\right]\] where \[\zsgn(x) = x
\sum_{i-1}^{\lfloor \zdeg/2\rfloor} \frac{\omega_i}{x^2 - \sigma_i^2}\] is the
Zolotarev approximation to the sign function, which is optimal in the
\(L_\infty\) norm over~\(\lowbound\leq|x| \leq1\).  The coefficients
\(\omega_i\) and \(\sigma_i\) can be computed in terms of Jacobi elliptic
functions\footnote{In particular \(\sigma_j=\lowbound\sn\Bigl(2i\K'(j-\half)
/\zdeg,\lowbound\Bigr)\), where \(\K'(k)=\K\left(\sqrt{1-k^2}\right)\) is a
complete elliptic integral.} depending upon~\(\lowbound\)
and~\(\zdeg\)~\cite{Zolotarev,vandenEshof:2002ms,kennedy:2003a,kennedy:2003b}.
Were the Zolotarev rational approximation expressed in a five dimensional
formalism, similar to that used for domain wall fermions, the size of the
fifth dimension would be~\(L_s=\zdeg\).

\begin{figure}
  \begin{center} \epspdffile{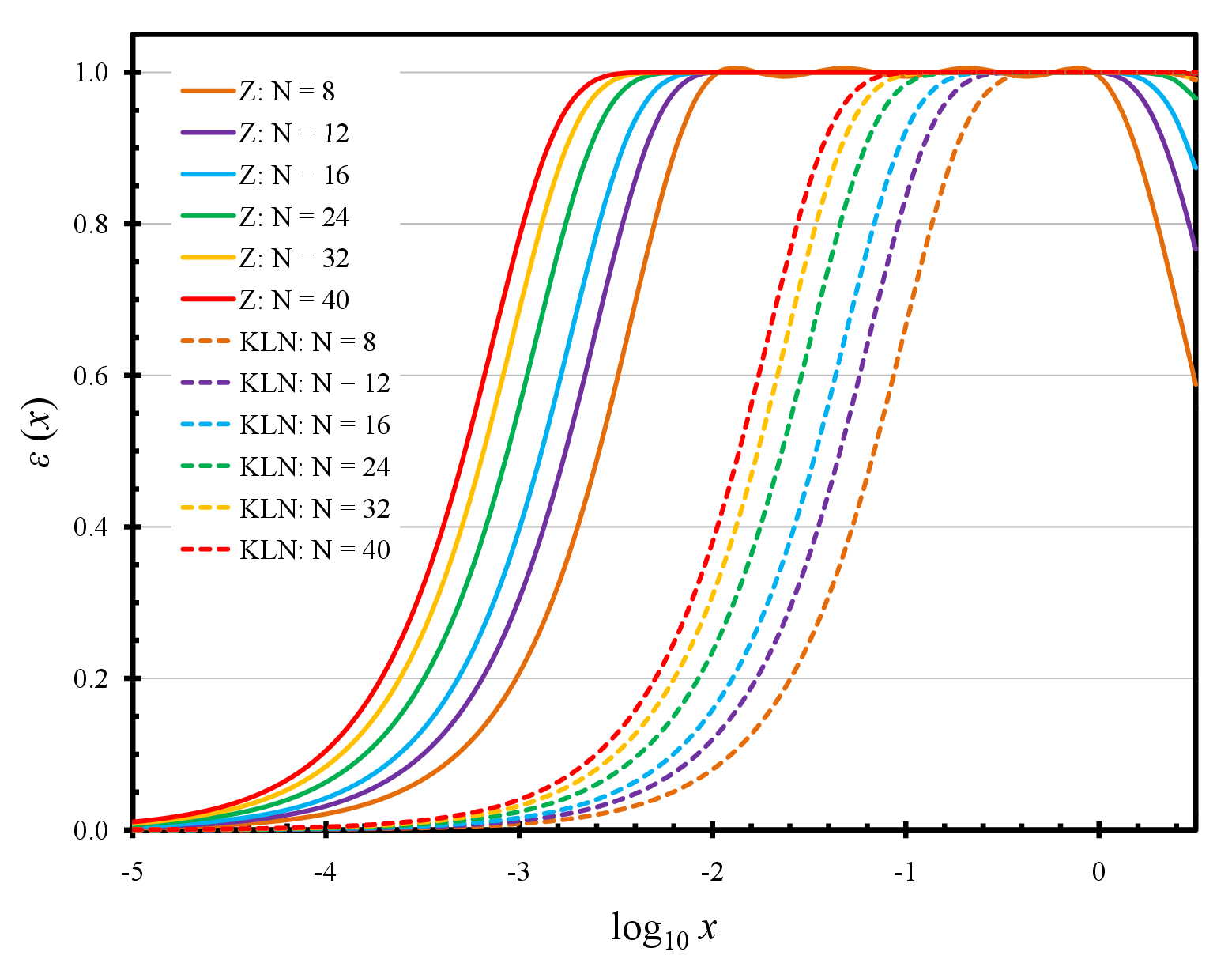} \end{center}
  \caption{The solid lines show Zolotarev rational approximations \(\zsgn(x)\)
    for various degrees \(\zdeg\) with \(\lowbound = 10^{-2}\) as a function
    of \(\log_{10}x\).  For comparison the dashed lines show the
    Kenney--Laub--Neuberger hyperbolic tangent approximations \(\hsgn(x)\) of
    the same degrees.}
  \label{fig:Zol-Tanh-lin-log}
\end{figure}

\begin{figure}
  \begin{center} \epspdffile{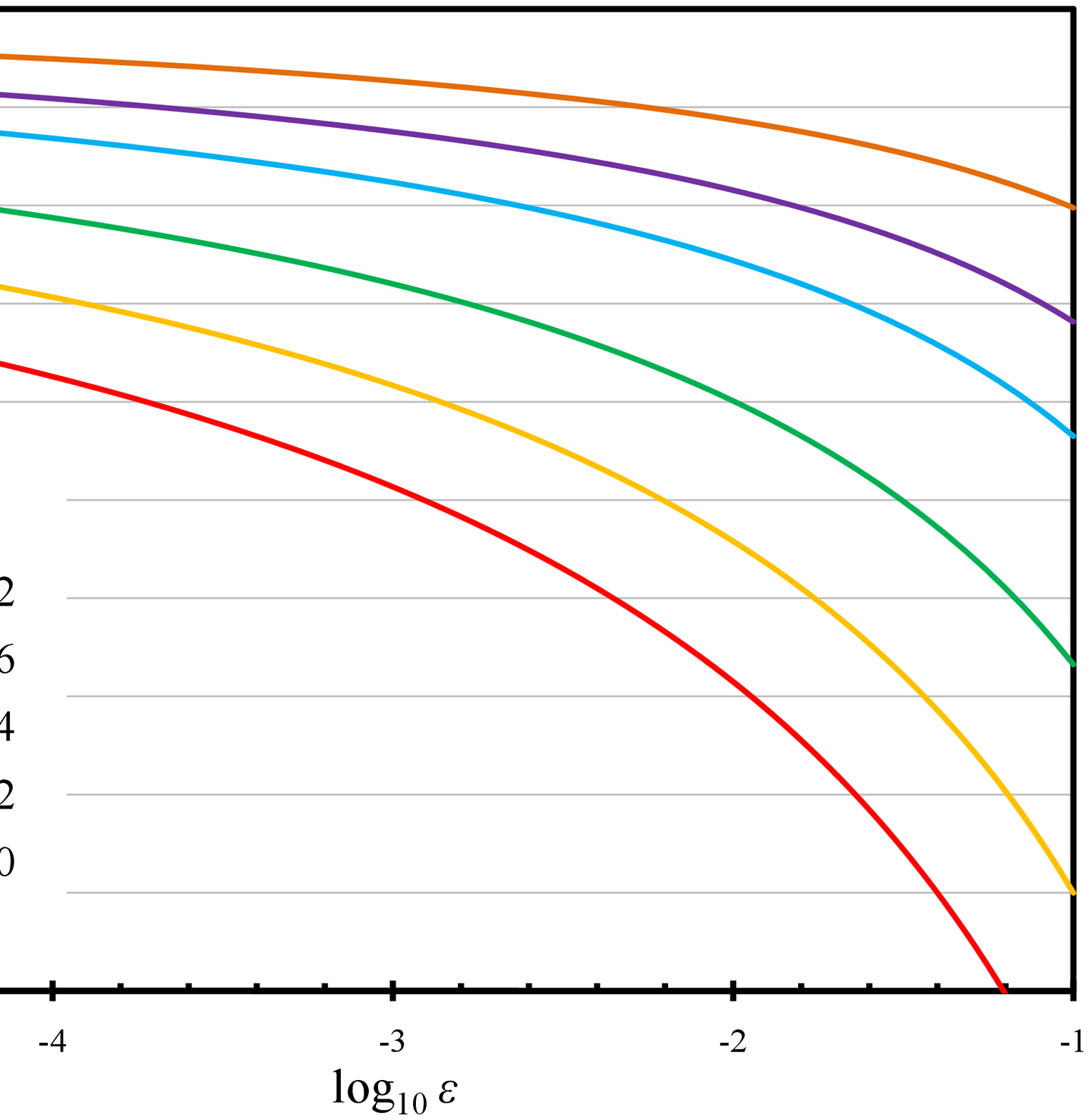} \end{center}
  \caption{The maximum error of the Zolotarev approximation \(Z(x)\) over its
    approximation interval, \(\displaystyle \Delta =
    \max_{\lowbound\leq|x|\leq1} |\zsgn(x)-1|\) for various degrees
    \(\zdeg\) are shown on a log-log plot.}
  \label{fig:Zol-err-log-log}
\end{figure}

There are two families of Zolotarev approximations, those that vanish at the
origin and those that are singular at the origin (up to a small re-scaling the
latter are just the reciprocals of the former).  We shall only use the
non-singular kind, which vary smoothly and monotonically from \(-1\) just
below \(-\lowbound\) to \(+1\) just above~\(\lowbound\).
Figure~\ref{fig:Zol-Tanh-lin-log} shows the Zolotarev approximation
\(\zsgn(x)\) for various degrees \(\zdeg\) and for a typical value
\(\lowbound=0.01\).
The maximum error of \(\zsgn\) over the interval on which it is optimal,
\(\displaystyle \Delta = \max_{\lowbound\leq|x|\leq1} |\zsgn(x)-1|\), is
shown in Figure~\ref{fig:Zol-err-log-log}.
Not only can we see from Figure~\ref{fig:Zol-err-log-log} that the error in
the approximation \(\Delta\) falls exponentially with the degree \(\zdeg\)
over the interval where the Zolotarev approximation is optimal, but also from
Figure~\ref{fig:Zol-Tanh-lin-log} we see that the approximation improves over
the interval~\(|x|<\lowbound\).

\subsection{The KLN Dirac operator}

The Kenney--Laub--Neuberger Dirac operator operator (KLN), which is implicitly
used in the domain wall approach
\cite{Kennedy:2006ax,Borici:2004pn,Chiu:2002ir}, is \[a\hop =
\half\left[(1+a\mu) + (1-a\mu)\gamma_5\, \hsgn\!\left(\frac H{\|H\|}
\right)\right]\] where the Kenney--Laub--Neuberger
\cite{kenney:1994,Neuberger:1998my} hyperbolic tangent approximation to the
sign function is \[\hsgn(x) = \tanh\left(\zdeg\tanh^{-1} x\right) =
\frac{(1+x)^\zdeg-(1-x)^\zdeg}{(1+x)^\zdeg+(1-x)^\zdeg}.\] This may be written
for even \(\zdeg\) as a partial fraction expansion~\cite{Kennedy:2006ax}
\[\hsgn(x) = \frac{2x}\zdeg \sum_{k=0}^{\frac{\zdeg}{2}-1}
\frac{1+\left(\tan\frac{(k+\half)
\pi}\zdeg\right)^2}{x^2+\left(\tan\frac{(k+\half)\pi}\zdeg\right)^2}.\]

Figures~\ref{fig:Zol-Tanh-lin-log} shows how this approximation compares to
the Zolotarev approximation.  Unlike the Zolotarev approximation it does not
require a minimum eigenvalue as an input; like the Zolotarev approximation the
accuracy depends on the order of the rational approximation.


\subsection{Fermionic forces in the Hybrid Monte Carlo algorithm}


One of the major possible difficulties with this method is that a very small
integration step size \(\dt\) may be required to prevent the molecular
dynamics (MD) trajectory in the Hybrid Monte Carlo (HMC) \cite{HMC} algorithm
from becoming unstable. For overlap fermions these instabilities can be
avoided (albeit with a little additional complexity in the force
calculation~\cite{Cundy:2007df}) because the small eigenvalues are not treated
by an approximation but deflated, and the simulation is run in the chiral
sector without zero modes~\cite{Bode:1999dd,Cundy:2005mr}. Here, because all
the eigenvalues are treated by the rational approximation, the force acting on
the gauge fields from their interaction with fermion fields might become large
for three reasons:
\begin{enumerate}
\item the fermion mass is small and the gauge-pseudofermion coupling involves
  the inverse of the Dirac operator which will, in general, have approximate
  zero modes;
\item the derivative of the chiral Dirac operator \(\nop\mu\) becomes large if
  the fermion field is close to a zero mode of the kernel operator~\(H\) (for
  an exact overlap operator, it is a Dirac \(\delta\)-function); and
  \label{item:dirac-step}
\item the estimate of the fermionic force obtained from pseudofermion fields is
  noisy~\cite{Egri:2005cx,Cundy:2008zc}. \label{item:pf-noise}
\end{enumerate}
The use of multiple pseudofermion fields \cite{Hasenbusch:2001ne,Clark:2006fx}
can reduce the effects of~(\ref{item:pf-noise}) and, for overlap fermions, a
factorisation of the determinant can can completely remove the effect of the
pseudofermion noise~\cite{Cundy:2008zc}.  In overlap simulations, various
transmission--reflection methods \cite{Fodor:2003bh,Cundy:2005pi,Cundy:2005mr}
have been introduced to resolve the Dirac \(\delta\)-function from the effects
of point \ref{item:dirac-step} above. However, the adaptations to the usual
HMC algorithm needed for exact overlap fermions are expensive, so for the case
of interest here --- where chiral symmetry is broken explicitly by a small
fermion mass \(a\mu\) --- we advocate choosing a Zolotarev approximation that
is not optimal for the smallest eigenvalues of \(H\), but is a compromise
between having a small residual mass \(\mres\ll\mu\) and having a small
fermionic force.  It is clear from Figure~\ref{fig:Zol-Tanh-lin-log} that we
may expect the Zolotarev Dirac operator to have a much smaller residual mass
than the KLN Dirac operator used in the domain wall method for any degree
\(\zdeg\) and reasonable values of~\(\lowbound\).

\begin{figure}
  \begin{center} \epspdffile{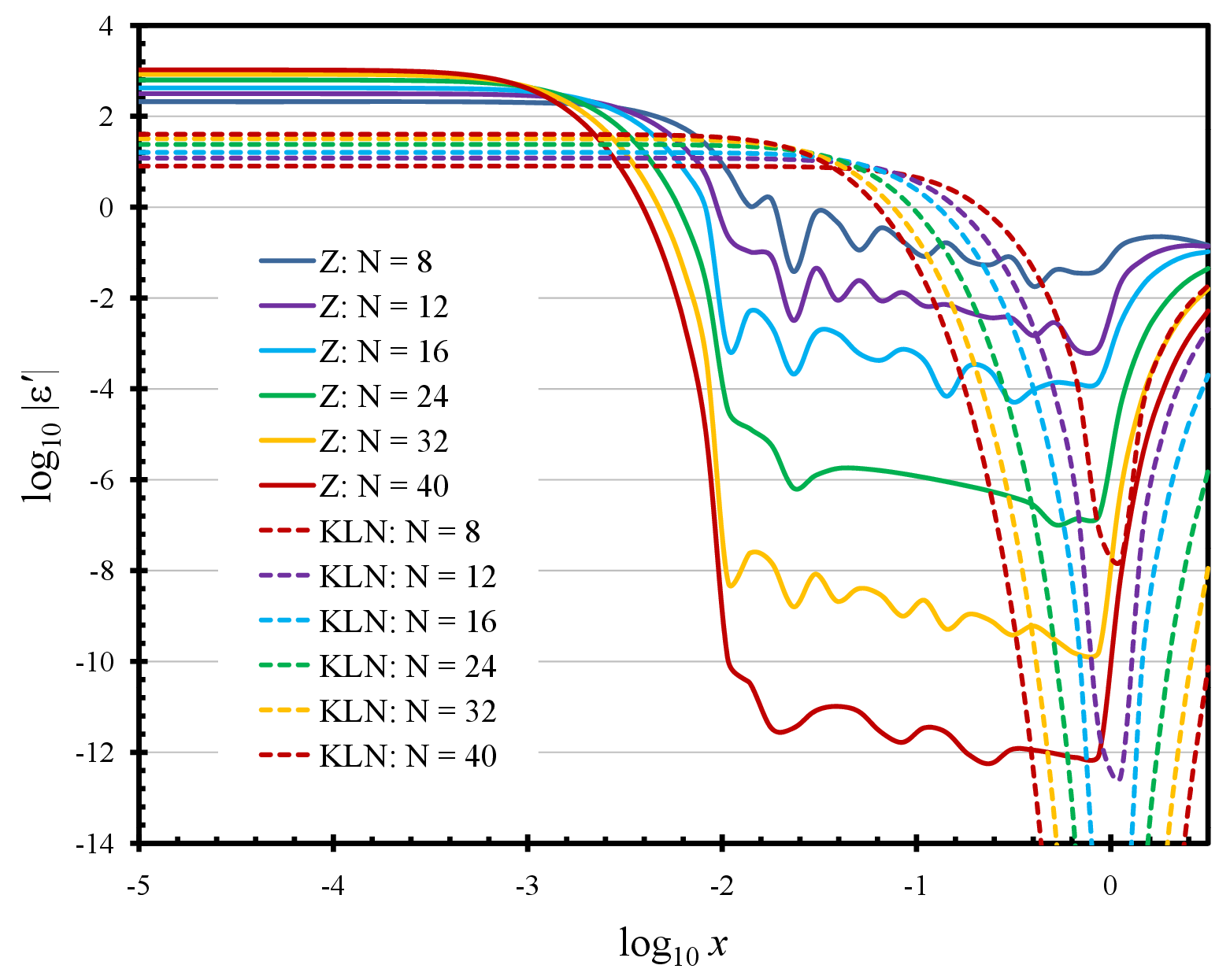} \end{center}
  \caption{The derivative of the Zolotarev rational approximation
    \(|d\zsgn(x)/dx|\) with \(\lowbound=10^{-2}\) and of the
    Kenney--Laub--Neuberger hyperbolic tangent approximation
    \(|d\hsgn(x)/dx|\) for various orders~\(\zdeg\).}
  \label{fig:Zol-Tanh-deriv-log-log}
\end{figure}

\begin{figure}
  \begin{center} \epspdffile{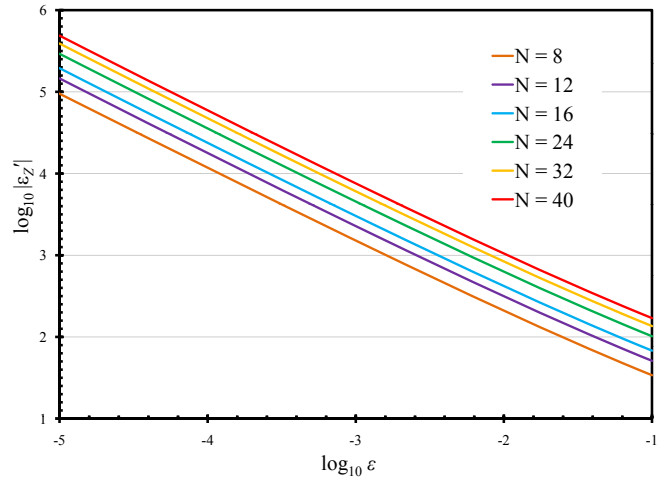} \end{center}
  \caption{The derivative of the Zolotarev rational approximation
    \(|d\zsgn/dx|\) at the origin \(x=0\) for various orders~\(\zdeg\).}
  \label{fig:Zol-deriv-origin-log-log}
\end{figure}

The fermionic force for continuous time MD evolution, to which the discrete
integrators used in HMC are a good approximation for reasonable acceptance
rates, contains a term proportional to the derivative of the approximation to
the sign function used in~\(\nop\mu\)~\cite{Cundy:2007df}.  The derivatives of
the Zolotarev and KLN approximations are shown in
Figure~\ref{fig:Zol-Tanh-deriv-log-log}, and several interesting features are
immediately obvious.
For both Zolotarev and KLN approximations the largest derivative occurs at
\(x=0\), this is trivial for the KLN approximation where the derivative at the
origin is \(\hsgn'(0)=\zdeg\), but it is less obvious for the Zolotarev
approximation.  We first note that the derivative \(\zsgn'(x)\) over the
interval \(\lowbound\leq|x|\leq1\) on which the approximation is optimal is
always small, and falls exponentially with the degree \(\zdeg\).  Next we see
that the derivative increases monotonically as \(x\) approaches zero where it
attains its maximum value.  In Figure~\ref{fig:Zol-deriv-origin-log-log} we
show the dependence of \(\zsgn'(0)\) on \(\lowbound\) and the degree~\(\zdeg\);
empirically the data are well fitted by
\(\zsgn'(0)\approx(0.49+0.025\zdeg)/\lowbound^{0.87}\).

\section{Numerical Implementation}  \label{sec:numerical-implementation}

\subsection{Markov Chains}

Our goal in this work is to investigate whether the Zolotarev Dirac operator
gives a smaller residual mass for the same cost as the domain wall Dirac
operator.  To do this, we have generated several small ensembles on
\(32^3\times64\) lattices starting from equilibrated domain wall
configurations.  We aimed to match the pion masses and lattice spacings for
these ensembles, although in practice the Zolotarev ensembles were at a
slightly lighter pion mass and slightly smaller lattice spacing.  We compared
the KLN Dirac operator, which is equivalent to a domain wall fermion with a
Bori\c{c}i kernel, with three different Zolotarev Dirac operators with
different values of~\(\lowbound\).  We used a Bori\c{c}i--Wilson kernel with
one step of over-improved~\cite{Moran:2008ra} stout
smearing~\cite{Morningstar:2003gk}, at \(\kappa=0.19\) and a tadpole improved
L\"uscher-Weisz gauge
action~\cite{Luscher:1984xn,Curci:1983an,Luscher:1985zq,Snippe:1997ru}.  We
ran enough trajectories in each case to ensure that the plaquette was
thermalised with the new action before taking measurements of the pion mass,
lattice spacing, and residual mass.  There is a small systematic uncertainty
in these measurements as our ensembles were not fully thermalized with respect
to the measured observables, however we have noticed no change in our
measurements along the Markov chains, so any systematic error due to
incomplete thermalization is probably smaller than our statistical errors.  We
have made measurements on 5--10 configurations for each run.  Our goal in
making these measurements is not to extract physics, but to obtain an
approximate idea of the physical parameters.

\subsection{MD Integrators}

Construction of the HMC MD integrator was straightforward.  On the larger
lattices, we used three additional pseudofermions, following the method of
Hasenbusch~\cite{Hasenbusch:2001ne} and multiple time-scale
integration~\cite{Sexton:1992nu} with the gauge field time steps being eight
times smaller than that for the the heaviest pseudofermion, which in turn had
a time step eight times smaller than that of the lightest pseudofermion.

\subsection{Linear Equation Solvers}

For any choice of rational approximation and kernel operator there are several
different approaches to the problem of inverting the chiral Dirac operator
\(\nop\mu\)~\cite{Kennedy:2006ax}.
\begin{enumerate}
\item Introduce a five dimensional matrix that has \(\nop\mu\) as its Schur
  complement.
  \begin{enumerate}
  \item Use this to find the inverse of \(\nop\mu\) applied to a four
    dimensional pseudofermion field in order to compute the fermionic force
    for four dimensional MD.
  \item Use the five dimensional matrix as part of a five dimensional MD
    scheme with five dimensional pseudofermions (this corresponds to the
    domain wall formalism).
  \end{enumerate}
  \item Apply the inverse of \(\nop\mu\) to a four dimensional pseudofermion
    field by use of a nested solver.
\end{enumerate}
In this paper we use a nested four dimensional solver, using a partial
fraction representation with relaxation~\cite{Arnold:2003sx}, GMRESR
preconditioning~\cite{Cundy:2004pza}, and deflation of about 70 kernel
eigenvalues.

It is not in the remit of this paper to compare the performance of four- and
five-dimensional solvers; our goal is to compare the effect of different
rational approximations. 
We expect that the relative performance of such different approximations to
the sign function should be similar with the use of five dimensional solvers
such as used in the domain wall algorithm.

The parameters of the runs are given in Table~\ref{tab:1}.  Pion masses were
measured using the pseudo-scalar correlator, and the lattice spacing using
\(r_0\)~\cite{Sommer:1993ce,Allton:2001sk}. Because we only have a handful of
configurations for each ensemble, it was difficult to get a reliable estimate
of the statistical errors, particularly for the lattice spacing, and a larger
study is required to get more accurate values. However, the physical pion
masses, which are in the range \(370-400\)~MeV, and the lattice spacings agree
across our ensembles to within the errors.
\begin{table}
  \begin{center}
    \begin{tabular}{cccccccc}
      \hline\hline
      Volume & \(\beta\) & \(10^4\lowbound \)  & \(\zdeg\) & \(a\) (fm)&
        \(am_\pi\) & \(k\) & Acc. \% \\
      \hline
      \(32^3\times64\) & \(8.65\) & KLN       & \(16\) & 0.101(3) & \(0.203(3)\)
        & ---    &  \(90\) \\
      \(32^3\times64\) & \(8.7\)  & \(1.93\)  & \(16\) & 0.094(4) & \(0.182(5)\)
        & \(35\) & \(100\) \\
      \(32^3\times64\) & \(8.7\)  & \(0.643\) & \(16\) & 0.096(3) & \(0.182(1)\)
        & \(18\) & \(97\)  \\
      \(32^3\times64\) & \(8.7\)  & \(0.193\) & \(16\) & 0.095(3) & \(0.183(2)\)
        & \(9\)  & \(92\)  \\
      \hline\hline
    \end{tabular}
  \end{center}
  \caption{Parameter values for ensembles. \(\beta\) is the L\"uscher-Weisz
    gauge coupling, \(a\) and \(m_\pi\) are the lattice spacing and pion mass
    respectively.  We also show \(\zdeg\), the degree of the rational
    approximation, and the HMC acceptance rate.  For the Zolotarev Dirac
    operator we list \(\lowbound\) and \(k\), the average number of
    eigenvalues of the kernel operator \(H\) that lie outside the optimal
    interval of the Zolotarev approximation.  The first line corresponds to
    our KLN approximation ensembles.}
  \label{tab:1}
\end{table}

\section{Results} \label{sec:results}

\subsection{HMC Acceptance Rate}

\begin{figure}
  \begin{center} \epspdffile{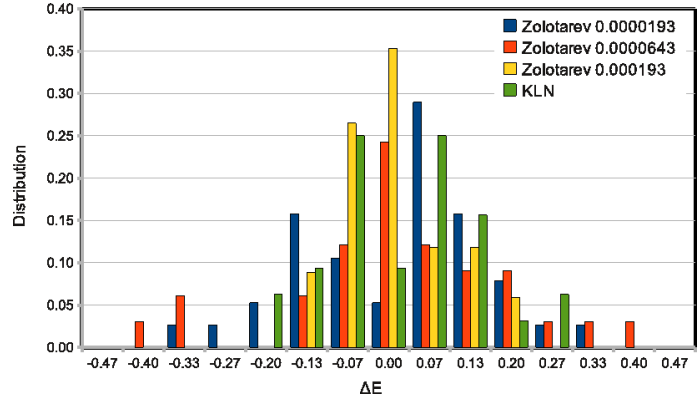} \end{center}
  \caption{The distribution of the HMC energy difference
    for the KLN and three Zolotarev Dirac operators for the
    \(32^3\times 64\) lattice.}
  \label{fig:zol7}
\end{figure}

The change in the value of the Hamiltonian \(\Delta E\) for our HMC runs is
 shown in Figure~\ref{fig:zol7}.  There are no large spikes, nor any large
 differences between the various different Dirac operators.  The acceptance
 rate shown in Tables~\ref{tab:1} are similar (albeit high) for each of the
 HMC runs.  The ensembles used the same Hasenbusch masses, MD time steps and,
 for our performance tests, the same number of deflated eigenvalues.  There is
 no indication of any MD instabilities.

\subsection{Performance}

The total time taken for the inversions required for each HMC run for the
\(32^3\times 64\) lattices is shown in Table~\ref{tab:HMCperformance}.  The
number of projected Wilson eigenvalues varied from trajectory to trajectory,
with an average around~\(70\).  All the measurements were performed with
identical time-steps, trajectory lengths, and an equal number of pseudofermion
fields.  While generating the configurations, the order of all the Zolotarev
Dirac and KLN operators was held fixed at~\(16\).
\begin{table}
  \begin{center}
    \begin{tabular}{ccccc}
      \hline\hline
      Volume & \(10^4\lowbound\) & Inverter & Eigenvalues \\
      \hline
      \(32^3\times64\) & KLN       & \(5,435\) & \(1,568\) \\
      \(32^3\times64\) & \(1.93\)  & \(6,291\) & \(1,627\) \\
      \(32^3\times64\) & \(0.643\)  & \(6,650\) & \(2,657\) \\
      \(32^3\times64\) & \(0.193\) & \(6,194\) & \(1,641\) \\
      \hline
    \end{tabular}
  \end{center}
  \caption{Time (in seconds) spent in the two dominant parts of the HMC code
    in each trajectory.}
  \label{tab:HMCperformance}
\end{table}  

While a larger study is needed for definitive results, it is clear that the
computational cost required for the Zolotarev and KLN runs does not differ by
any substantial amount.  The observed, small, difference is caused by a slower
convergence for the inversion of the Zolotarev Dirac operators compared to the
hyperbolic tangent operator, which is partially explained by the slightly
heavier quark mass for the KLN.  From these studies we conclude that the
difference in cost between running Zolotarev and KLN fermions at equal degree
and equal pion mass is at most about~20\%.  We must stress that these
inversions were not performed at equal residual mass; to run at the same
residual mass as the Zolotarev fermions, the cost for the KLN or domain wall
fermions would be far greater (see \S\ref{sec:chiralsymmetrybreaking}).

\subsection{Rate of topological charge changes}

\begin{figure} 
  \begin{center} \epspdffile{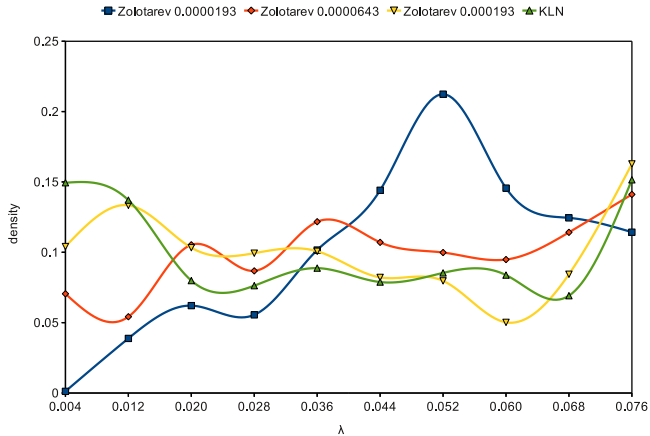} \end{center}
  \caption{The distribution of the smallest eight eigenvalues of the kernel operator
    \(H\) during the molecular dynamics for the KLN and Zolotarev Dirac
    operators.}
  \label{fig:eva}
\end{figure}

In Figure~\ref{fig:eva}, we plot the distribution of the ten smallest
eigenvalues of the kernel operator \(H\) as it evolved during the molecular
dynamics. From this limited data we cannot reliably estimate the rate of
topological charge change. The large fluctuations in the data are due to our small sample size; nonetheless the data is clear enough for a qualitative picture to emerge. 
Our data is consistent with our expectation that for the Zolotarev with
 largest \(\lowbound\) there is no suppression of the small kernel eigenvalues
 compared with KLN; although there are indications that the eigenvalues of the
 Zolotarev with the smallest \(\lowbound\) might be suppressed.  No difference
 can be seen from our data between the rate of topological tunnelling for the
 Zolotarev at large \(\lowbound\) and KLN Dirac operators.

The suppression of the small eigenvalues also aids the stability of the HMC
algorithm and the locality of the Zolotarev operator.

\subsection{Locality of Dirac operators}

We checked the locality of the Dirac operator by measuring the exponential
decay from a single source. All the Dirac operators are exponentially local,
and there is no noticeable difference in locality between the overlap operator
and the Dirac operators studied here.

\subsection{Chiral symmetry breaking}\label{sec:chiralsymmetrybreaking}

\begin{figure}
  \begin{center} \epspdffile{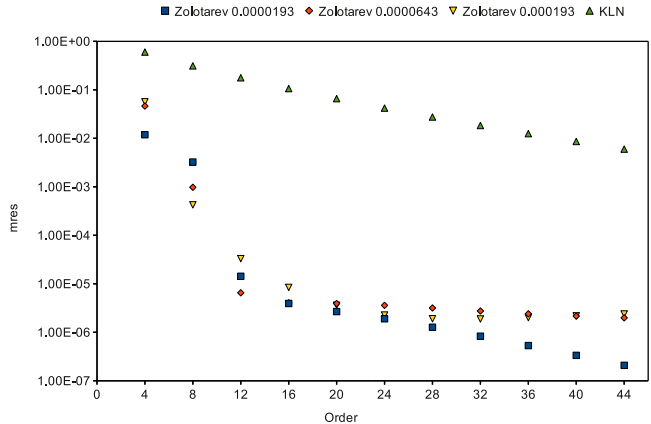} \end{center}
  \caption{The residual mass \(\mres\) plotted as a function of the degree
    \(\zdeg\) of the KLN and three Zolotarev approximations, as
    measured on the \(32^3\times 64\) lattices.}
  \label{fig:rm}
\end{figure}

The residual mass \(\mres\), as defined in \S\ref{sec:residual-mass}, is shown
for both the KLN and all three Zolotarev Dirac operators as a function of
the degree of the rational approximation in Figure~\ref{fig:rm}.  In this
plot, \(\mres\) has been averaged over all configurations.
It is immediately seen that both Zolotarev Dirac operators exhibit
significantly smaller residual masses than the KLN operator for all but the
lowest degrees.  At the order used in our tests, \(\zdeg = 16\), equivalent to
\(L_S=16\) for domain wall fermions, the improvement is four orders of
magnitude.  We also observe a striking difference in the way in which the
chiral symmetry is broken for the KLN and the Zolotarev Dirac operators.  For
the KLN Dirac operator the residual mass decreases steadily.  For the
Zolotarev Dirac operator it decreases rapidly up to around degree \(16\), and,
for the larger Zolotarev ranges reaches a plateau, where the fluctuations are
dominated by statistical noise, and for the smallest range continues to
decrease but at a slower rate as the order is increased.  Moreover, while
\(\mres\) is roughly constant between configurations for the KLN Dirac
operator, there is a large fluctuation between configurations for the
Zolotarev Dirac operator.  There is therefore little point in running a
Zolotarev approximation, at least for the parameter values considered here,
with degrees \(\zdeg>16\).

The pattern of chiral symmetry breaking for the Zolotarev Dirac operator can
be easily explained: there are two contributions to the violation of the
Ginsparg--Wilson relation for the Zolotarev Dirac operator, from the
imperfection of the approximation to the matrix sign function within the
interval on which the Zolotarev approximation is optimal (the ``bulk''), and
the contribution from the small eigenvalues below \(\lowbound\|H\|\).  The
first contribution decreases rapidly with \(\zdeg\), while the second
contribution decreases considerably more slowly, if at all.  For small
\(\zdeg\) \(\mres\) is dominated by the bulk, while for large \(\zdeg\) it is
dominated by the low modes.

\begin{figure}
  \begin{center} \epspdffile{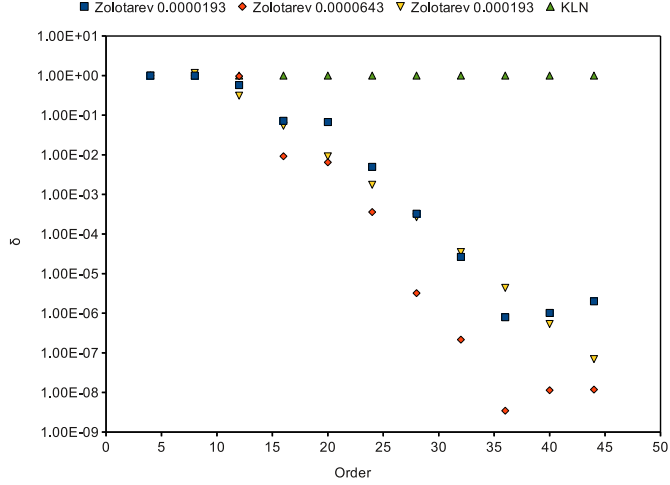} \end{center}
  \caption{The ratio of the contributions to \(\mres\) from bulk modes to that
    from low modes as a function of the degree of the rational function
    approximation \(\zdeg\) for the KLN and three Zolotarev Dirac
    operators, measured on the \(32^3\times 64\) lattices.}
  \label{fig:rm.lm}
\end{figure}
These two sources of explicit chiral symmetry violation are illustrated in
Figure~\ref{fig:rm.lm}, in which we have separated the contributions to
\(\mres\) from the bulk modes from the modes below \(\lowbound\|H\|\) and
plotted their ratio, defined as

\begin{equation}
  \delta = \frac{\sum_{x,x'} \left[{\nop\mu^\dagger}^{-1} 
      \gamma_5\defect\left(1-\sum_i |\psi_i\rangle\langle\psi_i|\right) 
      \nop\mu^{-1} \right]_{x,x'}}{\sum_{x,x'} \left[{\nop\mu^\dagger}^{-1} \gamma_5\defect 
      \left(\sum_i |\psi_i\rangle\langle\psi_i|\right)\nop\mu^{-1}
      \right]_{x,x'}},
\end{equation}
for for the lowest orthonormal eigenvectors, \(\psi_i\), of the kernel
operator \(H\).  It can be seen that for the Zolotarev approximation for order
\(\zdeg \ge 16\) the chiral symmetry breaking is almost entirely caused by the
lowest eigenvalues of the kernel operator, while for lower \(\zdeg\) the bulk
eigenvalues have a larger contribution.  It is thus clear that the plateau in
\(\mres\) for the \(\lowbound\|H\|=0.0000643\) ensemble is caused by the low
modes, whereas the \(\lowbound\|H\|=0.0000193\) ensemble has no such small
eigenvalues, and correspondingly the residual mass gradually improves as the
order of the rational approximation improves and the approximation for the
matrix sign function gets better for the bulk modes.

\section{Conclusions} \label{sec:conclusions}

Our main conclusion is that the Zolotarev Dirac operator seems to provide a
significant improvement over both domain wall and overlap fermions for
computations with light fermions on fine lattices.

On coarse lattices contributions from the small eigenvalues subspace of the
kernel operator dominate the residual mass.  Some of these eigenvectors are
topological, whereas others are lattice artefacts (``defects'').  Various
mechanisms have been suggested to suppress such defects without affecting the
underlying continuum physics by increasing autocorrelations for topology
change.  To the extent that such methods are effective the contributions to
the residual mass from the bulk will still be significantly reduced by the
Zolotarev Dirac operator.

\subsection{Comparison with Domain Wall fermions}

The principal advantage that the Zolotarev Dirac operator has over domain wall
fermions is that it gives a significantly smaller residual mass for the same
computational effort.  In particular, this allows for simulations at smaller
lattice spacing than are currently possible with domain wall fermions.  The
second advantage is that the Zolotarev Dirac operator can be tuned to balance
the performance of the HMC, the tunnelling rate and the residual mass.

There are still questions which need to be addressed in future work. In particular, we have not addressed the question of whether using five dimensional (as used by Chiu and collaborators~\cite{Chiu:2002ir}) or four dimensional pseudofermions as in this work is superior. Should the five dimensional inversion prove superior to the nested four dimensional inversion (the comparison in~\cite{Hashimoto:2007vv} used a considerably sub-optimal nested 4D algorithm, so this question remains open), then this can easily incorporated into the four dimensional algorithm, so it seems unlikely that there is much difference between the two. However, until a direct comparison is made in a future work, no definite statement can be made in favour of either formulation.

\subsection{Comparison with Overlap fermions}

The Zolotarev Dirac operator has the advantage over overlap fermions that it
is faster.  An exact overlap calculation requires an accurate resolution of
the matrix sign function during the molecular dynamics, which in practice
requires the transmission/reflection
algorithm~\cite{Fodor:2003bh,Cundy:2005pi,Cundy:2005mr}. To allow frequent
topological charge changes, the overlap HMC algorithm has to be further
refined and carefully tuned~\cite{Cundy:2008zc}, leading to approximately a
doubling of the cost per trajectory. Furthermore, while these refinements
allow topological charge changes every few trajectories, there may still be
longer autocorrelations than with the Zolotarev Dirac operator.  This cost can
be removed by adding unphysical terms to the action~\cite{Hashimoto:2006rb},
thus forbidding both topological tunnelling and kernel eigenvalues close to
zero. The effect of this with regards to possible artefacts and ergodicity is,
however, unclear.

\section*{Acknowledgements}
The work has been performed under the HPC-EUROPA2 project (project number: 228398) with the support of the European Commission - Capacities Area - Research Infrastructures.
We would like to thank the Edinburgh Parallel Computing Center for supporting and hosting NC's visit to Edinburgh. Simulations were run on the
Blue Gene/P at the J\"ulich Supercomputing Center, and the supercomputer
Hector at the EPCC. We are thankful for the help and comments of Chris Johnson, Chris
Maynard, Thomas Lippert, Stefan Krieg, Giannis Koutsou, Peter Boyle, and Ting Wai Chiu. Initial configurations were taken from the RBC/UKQCD domain wall ensembles~\cite{Antonio:2006px}. NC
was supported by funding from the DFB SFB TR55 and the BK21 program funded by NRF, Republic of Korea. Some of this work was carried out at the workshop `Perspectives and challenges for full QCD lattice
calculations' from 5-9 May 2008 at the ECT*, Trento and the workshop `Lattice
Quantum Chromodynamics' at the Kavli Institute for Theoretical Physics, during
July 2009 in Beijing.

\raggedright
\bibliographystyle{elsarticle-num}
\bibliography{zol-dirac-op}

\end{document}